\documentclass[10pt,conference]{IEEEtran}
\IEEEoverridecommandlockouts
\usepackage{cite}
\usepackage[table]{xcolor}
\usepackage{multirow}
\usepackage{tabularx}
\usepackage{amsmath,amssymb,amsfonts}
\usepackage{algorithmic}
\usepackage{graphicx}
\usepackage{textcomp}
\usepackage{xcolor}
\usepackage{todonotes}
\usepackage[linesnumbered,lined,boxed,commentsnumbered]{algorithm2e}
\usepackage{caption}
\usepackage{subcaption}

\def\BibTeX{{\rm B\kern-.05em{\sc i\kern-.025em b}\kern-.08em
    T\kern-.1667em\lower.7ex\hbox{E}\kern-.125emX}}
\begin{document}

\title{Accelerating High-Order Stencils on GPUs}

\author{\IEEEauthorblockN{Ryuichi Sai}
\IEEEauthorblockA{\textit{Department of Computer Science} \\
\textit{Rice University}\\
Houston, TX, USA \\
ryuichi@rice.edu}
\and
\IEEEauthorblockN{John Mellor-Crummey}
\IEEEauthorblockA{\textit{Department of Computer Science} \\
\textit{Rice University}\\
Houston, TX, USA \\
johnmc@rice.edu}
\and
\IEEEauthorblockN{Xiaozhu Meng}
\IEEEauthorblockA{\textit{Department of Computer Science} \\
\textit{Rice University}\\
Houston, TX, USA \\
xm13@rice.edu}
\and
\IEEEauthorblockN{Mauricio Araya-Polo}
\IEEEauthorblockA{\textit{Computational Science and Engineering} \\
\textit{Total E\&P Research and Technology US, LLC.}\\
Houston, TX, USA}
\and
\IEEEauthorblockN{Jie Meng}
\IEEEauthorblockA{\textit{Computational Science and Engineering} \\
\textit{Total E\&P Research and Technology US, LLC.}\\
Houston, TX, USA}
}

\maketitle

\begin{abstract}
Stencil computations are widely used in HPC applications. Today, many HPC platforms use GPUs as accelerators. As a result, understanding how to perform stencil computations fast on GPUs is important.
While implementation strategies for low-order stencils on GPUs have been well-studied in the literature,
not all of the techniques work well for high-order stencils, such as those used for seismic imaging.
Furthermore, coping with boundary conditions often requires different computational logic, which complicates efficient exploitation of the thread-level parallelism on GPUs.
In this paper,
we study practical seismic imaging computations on GPUs using high-order stencils on large domains with meaningful boundary conditions.
We manually crafted a collection of implementations of a 25-point seismic modeling stencil in CUDA along with code to apply the boundary conditions. We evaluated our stencil code shapes,
        memory hierarchy usage,
        data-fetching patterns,
        and  other performance attributes.
We conducted an empirical evaluation of these stencils using several mature and emerging tools and discuss our quantitative findings.
Among our implementations, we achieve twice the performance of a proprietary code developed in C and mapped to GPUs using OpenACC.
Additionally, several of our implementations
    have excellent performance portability.
\end{abstract}


\section{Introduction}

"Remember that Time is Money." \cite{fb1748}
This is even more true in today's competitive business environments,
    such as the oil and gas industry,
    where fast simulations enable more realizations of experiments that can reduce the uncertainty of hydrocarbon reserves location or CO$_{2}$ storage management processes.
Seismic depth imaging is the main tool used to extract information
    from seismic field records to identify relevant subsurface structures. High-order stencil computations typically serve as the foundation for seismic depth imaging.
Over the past two decades, the availability of sufficiently powerful computational resources has led to the every-day use of more complex stencil-based wave equation approximations, and the power of today's petascale systems enables simulations based on the full-wave equation instead of simple approximations.

Today, HPC platforms often employ Graphics Processing Units (GPUs) to increase their computational power. Accordingly, using GPUs to accelerate full-wave equation simulations based on high-order stencils is a natural approach.
However, the complexity of GPU architectures makes achieving top performance with high-order stencil computations  surprisingly difficult.
Without careful design, a stencil computation on a GPU is likely to underperform. An efficient implementation of high-order stencils with boundary conditions on a GPU requires paying careful attention to data reuse, warp utilization, work balance, and arithmetic intensity among other issues. For that reason, understanding how to develop efficient high-order stencils for GPUs is a topic of significant interest.

Since GPUs from different vendors have different characteristics and the characteristics of GPUs from a single vendor often change significantly between generations, 
performance portability across GPUs with varying characteristics is of significant interest. The best kernel on one GPU may not be the best on GPUs from other vendors and may not remain the best on newer generations of GPUs.


For these reasons, our current goal is to identify how to 
    achieve excellent performance for high-order stencils on GPUs, and understand the factors that affect performance portability.
%
To do so,
    we need to understand the strengths and weaknesses of various code shapes for high-order stencils.
This paper describes our progress toward this goal and makes the following contributions:

\begin{itemize}
    \item a careful comparison of existing approaches,
            including an assessment of their strengths and weaknesses when
            applied to high-order stencils with boundary conditions;
    \item the implementation and tuning of 
            a collection of high-order stencil kernels
            with a selected set of algorithms and their variants
            using CUDA;
    \item a performance comparison of stencil implementations across multiple generations of NVIDIA GPUs along with a quantitative assessment of their performance using a Roofline performance model for GPUs; and
    \item an investigation of the characteristics of different stencil kernels that affect their performance
\end{itemize}

The next section is an overview of related work. Sections \ref{sec:approach} and \ref{sec:impl} present our approaches and implementations, respectively.
Section \ref{sec:evaluation} describes our evaluation methodology, experimental results, and a discussion of our findings.
Section~\ref{sec:concl} summarizes our conclusions and briefly discusses our plans for future work.

\section{Related Work}
\label{sec:related}
There is a rich literature describing efforts to efficiently implement stencil computations on CPUs~\cite{cache-oblivious-1},
    \cite{cache-oblivious-2},
    \cite{cache-oblivious-3},
    \cite{cache-oblivious-4},
    \cite{cache-oblivious-5},
    \cite{time-skewing-4},
    \cite{time-skewing-5},
    \cite{time-skewing-1},
    \cite{time-skewing-2},
    \cite{time-skewing-3},
    \cite{semi1},
    \cite{semi2},
    \cite{krishnamoorthy}
    and GPUs~\cite{holewinski-overlapped},
    \cite{grosser-split},
    \cite{nguyen},
    \cite{micike},
    \cite{matsu},
    \cite{rawat-sdslc},
    \cite{rawat-dso}.
We discuss the most related efforts below.

Time skewing~\cite{time-skewing-4},
\cite{time-skewing-5} accelerates stencil computations by increasing data reuse and cache locality by skewing one or more data dimensions by the time dimension so that several time steps can be computed for a tile while values are in cache. It has been widely used on CPUs, e.g., 
\cite{time-skewing-1},
\cite{time-skewing-2}, and
\cite{time-skewing-3}.


Overlapped tiling uses time skewing 
    to increase the arithmetic intensity of parallel stencil computations by trading redundant computation along the boundaries of overlapped tiles for a reduction in memory bandwidth required \cite{krishnamoorthy}, \cite{holewinski-overlapped}.
Overlapped tiling is effective on GPUs because loading data from a GPU's global memory is much more costly than data-parallel computation. Furthermore, redundant computation can be overlapped with data accesses to help hide memory latency.
While overlapped tiling has been shown to improve the performance of low-order stencils on GPUs, for high-order stencils, redundant computation grows quickly when skewed across multiple time steps by the width of a high-order stencil. 

Split tiling \cite{grosser-split} is an alternate approach for accelerating computation with time skewing. Rather than using overlapped tiles, which can introduce large amounts of redundant computation, split tiling computes points in two phases.
The first phase computes tiles in parallel as hypertrapezoids that taper along the time dimension.
Once all tiles from the first phase have been computed, a second phase back-fills the missing points in the time dimension.

The semi-stencil algorithm \cite{semi1}, \cite{semi2}, which has only been studied on CPUs, factors the computation of a stencil into two or more pieces. Rather than computing the result for a point as a single computation, the semi-stencil algorithm divides computation along one or more axes into two halves. While sweeping along a dimension, it computes the final result for one point and a partial result for another point a half-stencil width ahead. Compared to the traditional implementation of stencils, this approach changes the load/store ratio for the computation
    by trading half of the loads along a dimension for a store and a reload of a partial result. On a GPU, this has the potential for nearly halving the cache footprint of a thread block for a high-order stencil.

Nguyen et al. \cite{nguyen} introduce a 3.5D blocking algorithm as
    a mix of 2.5D spatial blocking with 1D temporal blocking.
2.5D spatial blocking involves
     blocking in a 2D plane
    and streaming along a third dimension.
To increase data reuse, they store active 2D planes in GPU shared memory. In a 3.5D variant, they employ time skewing to advance the computation for multiple time steps before writing data back to the global memory.
While the 3.5D algorithm works very well on CPUs,
    the 1D temporal blocking
    introduces two potential implementation challenges
        for high-order stencils with boundary conditions on GPUs:
 barrier synchronizations and limited parallelism.
In this paper, we evaluate  2.5D spatial blocking of high-order stencils and plan to explore 3.5D blocking in future work.

Nguyen et al.'s approach \cite{nguyen} loads a central  plane along with halo planes above and below 
        into shared memory
    for faster data access while computing stencil operations for points in the central plane.
While this strategy improves data reuse,
    the size of a data tile is limited
        by the GPU shared memory size.
To reduce the shared memory pressure,
    we looked into the work that uses registers on GPUs.
Micikevicius \cite{micike} also uses 2.5D blocking; 
however, his approach maintains data points along the third dimension in registers rather than in shared memory.

Recently, as part of their AN5D framework work,
    Matsumura et al. \cite{matsu} apply three more refinements to
    2.5D and 3.5D solutions:
    fixed register allocations,
    double buffering,
    and division of the streaming dimension.
While these approaches work extremely well for simple single-statement kernels,
    neither boundary conditions nor multi-statement stencils
        are evaluated.
In our work, we study a high-order stencil with boundary conditions,
    and part of our application has multiple statements,
    instead of simple single-statement stencil updates.

Other interesting approaches to tackle
    the stencil computations 
    including
    auto-tuning with dynamic resource allocations \cite{rawat-autotuning},
    DAG reordering \cite{rawat-reordering},
    diamond tiling using polyhedral model \cite{uday}, \cite{uday-student},
    functional programming \cite{lift}, \cite{rise},
    and multi-layer intermediate representations \cite{mlir1}, \cite{mlir2}, \cite{mlir3}.

From a software engineering perspective,
    there are two strategies for developing stencils for NVIDIA GPUs:
    hand-written kernels in CUDA
    and Domain-Specific Language (DSL)-based approaches
    \cite{cache-oblivious-5},
    \cite{rawat-sdslc},
    \cite{dsl1},
    \cite{dsl2},
    \cite{dsl3},
    \cite{dsl4},
    \cite{dsl5}.
The DSL approach can simplify the generation of code with complex logic.
While we are interested in DSL-based approaches for the future,
    the focus of this paper is to understand in  detail the strengths and weaknesses of various algorithmic strategies for achieving high performance and performance portability for high-order stencils. To avoid limitations as we explore this space, we chose to evaluate hand-written kernels.

\section{Approach}
\label{sec:approach}

We developed several implementations of the acoustic isotropic approximation of the wave equation~\cite{total-minimod} used for seismic imaging by the oil and gas industry.
Solving this with finite differences involves using a high-order
stencil-based solver with suitable boundary conditions.
Oil and gas applications use such strategies 
on large grids to model subsurface and generate seismic data from source perturbations.
In our work, we employ different code shapes that differ principally in how they organize the computation (e.g., 2D vs. 3D tiles) and how they manage the memory hierarchy.
In the rest of this section,
we will briefly describe the acoustic isotropic approximation,
explain the various data decomposition strategies we employ, describe blocking strategies, and discuss how we structure our implementations.

\subsection{Seismic Modeling and Acoustic Isotropic Kernel}
\label{sec:seismic-modeling-acoustic-iso}

We study a stencil-based implementation
    of the acoustic isotropic wave equation approximation for seismic modeling.
The details of the model are described in \cite{total-minimod}. The wave equation  for an acoustic isotropic operator with constant-density has the following form:
\begin{equation}
\frac{1}{\mathbf{V}^2}\frac{\partial^2 \mathbf{u}}{\partial t^2} - \nabla^2 \mathbf{u} = \mathbf{f},
\end{equation}
where $\mathbf{u} = \mathbf{u}(x,y,z)$ is the wavefield, $\mathbf{V}$ is the Earth model (with velocity as rock property), and $\mathbf{f}$ is the source perturbation. The equation is discretized in time using a second-order centered stencil, resulting in the semi-discretized equation:
\begin{equation}
\mathbf{u}^{n+1} - \mathbf{Q}\mathbf{u}^n + \mathbf{u}^{n-1} = \left(\Delta t^2\right) \mathbf{V}^2 \mathbf{f}^n,\label{eq:minimod-semidisc}
	\mathrm{with\ }\mathbf{Q} = 2 + \Delta t^2 \mathbf{V}^2 \nabla^2.
\end{equation}
Finally, the equation is discretized in space using a 25-point stencil in 3D, with eight points in along each axis surrounding a center point:

\begin{multline}
\nabla^2 \mathbf{u}(x,y,z) \approx 
c_{xyz} \times \mathbf{u}(i,j,k) + \\
\mspace{65mu} \sum_{m=1}^4 c_{xm}\times\left[\mathbf{u}(i+m,j,k) + \mathbf{u}(i-m,j,k)\right] + \\
\mspace{105mu} c_{ym}\times\left[\mathbf{u}(i,j+m,k) + \mathbf{u}(i,j-m,k)\right] + \\
c_{zm}\times\left[\mathbf{u}(i,j,k+m) + \mathbf{u}(i,j,k-m)\right]
\end{multline}

where $c_{xyz}, c_{xm}, c_{ym}, c_{zm}$ are the discretization parameters.

A high-level description of the algorithm is shown in Algorithm \ref{algo:minimod}. We apply a Perfectly-Matched Layer (PML) \cite{PML} boundary condition to the regions around the physical domain. The resulting extended domain consists of an ``inner'' region where Equation (\ref{eq:minimod-semidisc}) is applied, and the outer ``boundary'' region where a PML calculation is applied.

\begin{algorithm}[t]
	\KwData{$\mathbf{f}$: source}
	\KwResult{$\mathbf{u}^n$: wavefield at timestep $n$, for $n\leftarrow 1$ \KwTo $T$}
	$\mathbf{u}^0 := 0$\;
	\For{$n\leftarrow 1$ \KwTo $T$}{\nllabel{line:tsloop}
		\For{each point in wavefield $\mathbf{u}^n$}{
			Solve Eq.~\ref{eq:minimod-semidisc} (left hand side) for wavefield $\mathbf{u}^n$\;
		}
		$\mathbf{u}^n = \mathbf{u}^n + \mathbf{f}^n$ (Eq.~\ref{eq:minimod-semidisc} right hand side)\;
	}\nllabel{line:end-of-ts}
	\caption{A high-level description of the algorithm for solving the acoustic isotropic approximation of the wave equation with constant density.}
	\label{algo:minimod}
\end{algorithm}

To solve the acoustic isotropic approximation for the wave equation, in the inner region we apply a  
    complex multi-statement stencil 
    that is 8th-order in space and 2nd-order in time.
This involves applying
    a star shaped 25-point stencil to data stored in the $u$-array.
In the PML layer, we employ a 7-point star shaped stencil to compute boundary conditions. 
The data for this 7-point stencil is stored in the $eta$-array.

In production simulations,
    the grid (that represent the physical domain) size is usually large (up to $4,000$ grid points in each dimension).
To yield results of practical use, the stencil computations need to be applied iteratively for
a large number of time steps.

While we specifically study the
acoustic isotropic kernel as the seismic model wave approximation in this paper,
we believe that our approach is general enough
        that it could be applied to
        other high-order stencils with boundary conditions.

\subsection{Data Domain Decomposition}

As described in the previous section,
    our data domain contains two regions,
    the inner region
    and the Perfectly Matched Layer (PML) boundaries.
The inner region is a cubic grid sits at the center of the data domain and
     the PML region represents the volume between the inner region and the data domain boundaries.
The size of the inner region and the width of the PML region
    are defined as inputs to a simulation.

GPUs have different architectural characteristics than CPUs. As a result, GPU computations must be structured differently than CPU computations to achieve high performance. First, the Single-Instruction-Multiple-Thread (SIMT) execution model used by GPUs differs significantly from the execution model on  CPUs. On NVIDIA GPUs, the execution model is realized by scheduling groups of  32 SIMT threads known as warps. To exploit thread-level parallelism in the SIMT model, GPU computations must utilize fine-grain data parallelism. 
A group of warps constitute a block which has its own quota for shared memory,
    registers, and other hardware resources; the number of blocks that can be active simultaneously is limited by the aggregated resource quota of active threads enforced by hardware limits.
Second, since GPUs have a memory hierarchy distinct from CPUs, computations must be appropriately structured to exploit the GPU memory hierarchy.
Finally, on NVIDIA GPUs, one could realize coarse-grain parallelism across Streaming Multiprocessors (SMs) by partitioning the computation into a sufficient number of blocks to keep the SMs busy.

We experimented with three decomposition strategies
    based on our data domain and boundary conditions.

First, we developed a single kernel that could be applied to any region of the data domain. The kernel contains conditionals that employ the PML calculations near any of the domain boundaries and compute the stencil for the acoustic isotropic wave function approximation in the inner region of the data domain.
This strategy yields  branch divergence for subregions that contain points in both the PML and inner regions which hurt performance.

Next, we developed separate kernels for the central region and the PML region.
These separate kernels can be launched concurrently.
This strategy eliminates the need for checking whether the point is inside inner region or PML region in every kernel,
    thus, reducing the chance of branch divergence.
    Nevertheless, it leaves unbalanced work among threads,
    along the boundaries between the inner and PML regions when the size of the GPU blocks doesn't evenly divide the extents of the PML and central regions.

\begin{figure}[t]
    \centering
    \includegraphics[width=0.48\textwidth]{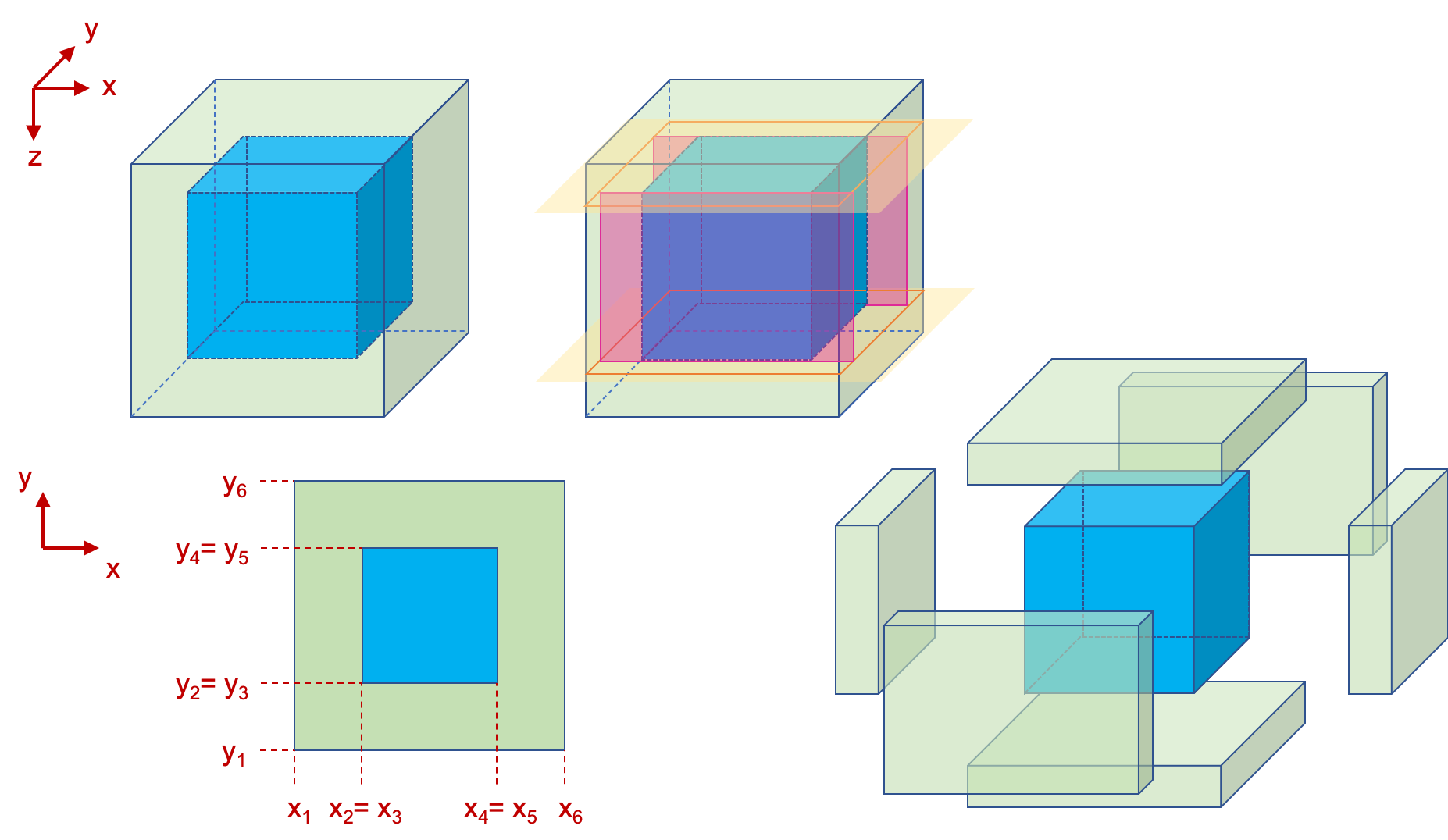}
    \caption{Data Domain Decomposition}
    \label{figure:data-decomposition}
\end{figure}

Lastly, we developed the strategy as shown in Figure \ref{figure:data-decomposition}.
We separate the inner region from PML region,
    and further divide PML region into six subregions.
We slice the domain along the top and bottom of the inner region,
    and this gives us a top block, a bottom block,
    and a border of four walls.
We further slice along the front and back,
    and it becomes four separate walls.
These four walls and the two subregions from the first cuts result in
    total of six subregions of the PML region,
    namely: top, bottom, front, back, left and right subregions.
The symmetry of these subregions is a relevant characteristic that we discuss later
    along with our results.
Next, we launch individual GPU kernels of stencil computations for each of the seven subregions:
    one for the inner region and six for the PML subregions.
This approach does not have intrinsic branch divergence at the boundaries.
While there are still work imbalances due to different grid sizes,
    they occur only for a few edge cases along the borders.
We could further reduce unbalanced work by using
automated code generation that tailors the number of threads to match the number of points at border locations.

\subsection{Blocking Strategies}

For each of the seven regions,
    we further slice it into smaller blocks,
    so that each block can fit into
        GPU's resources for each kernel launch.
We use two blocking strategies in our experiments:
    3D Blocking and 2.5D Blocking.

\begin{figure}[t]
    \centering
    \includegraphics[width=0.48\textwidth]{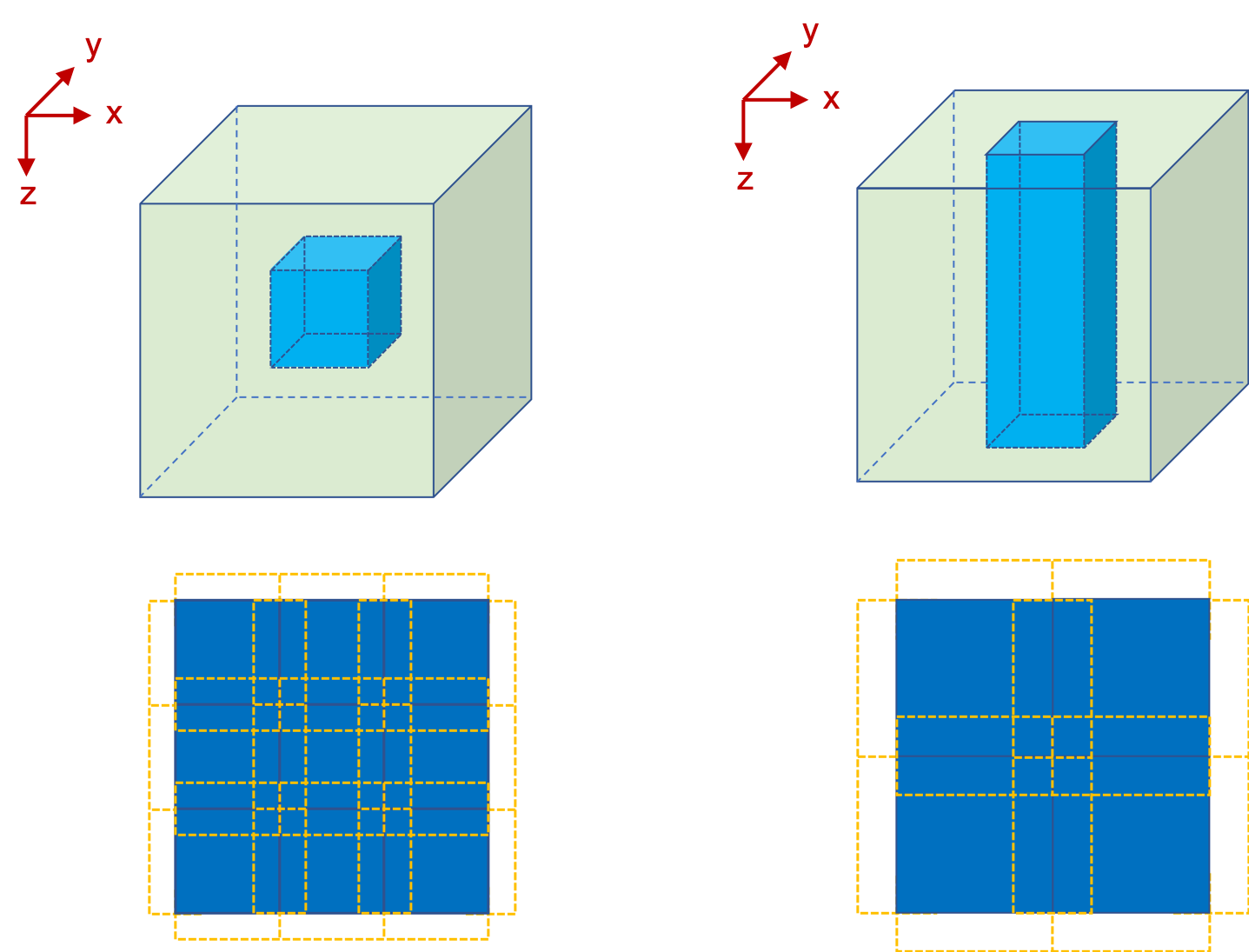}
    \caption{Blocking Strategies: (left) 3D Blocking; (right) 2.5D Blocking}
    \label{figure:blocking-strategies}
\end{figure}

\subsubsection{3D Blocking}

We divide each of the data regions
    into axis-aligned 3D blocks.
To find the best block dimensions,
   we use fixed values in each execution to simplify experiments with different values.
To perform stencil computations on GPUs,
    each block maps to a kernel launch with a 3D thread block
        with the thread dimensions matching the block dimensions.
All points inside the block and their halos
    are explicitly copied into the GPU on-chip memory before any kernel is launched.

\subsubsection{2.5D Blocking}

We partition the data domain along the inner two X and Y data dimensions and perform a streaming computation along the outermost Z dimension.
We launch kernels with 2D thread blocks with their dimensions matching the 2D planes.

\subsection{Kernels and their variants}

We implemented several kernels.
Each kernel employs a different combination of strategies for  blocking, managing data accesses, and traversing the data volume.
To better understand the strengths and weaknesses of these implementation alternatives,  
    every implementation has multiple variants,
    which employ different tile sizes.
We describe the details of our kernel implementations in the next section.

\section{Implementations}
\label{sec:impl}

In describing our implementations, we use
${R}$ to denote the width of the halo,
    which is half the spatial order of the stencil.
For the acoustic isotropic simulations in our experiments,
    ${R}$ is ${4}$.
Let ${Nx}$, ${Ny}$, and ${Nz}$ denote the extents
    of the input data region along the ${X}$, ${Y}$, and ${Z}$ axes,
    respectively.
For 3D blocks,
    we use ${(x, y, z)}$ to denote the 3D coordinate for
        both a point location in a 3D block
        and the thread in a kernel thread block.
Similarly, for 2D planes,
    ${(x, y)}$ is used to locate a point in the 2D plane,
    as well as identifying the thread.

\subsubsection{3D Blocking Using Global Memory Only}
This is conceptually and practically the simplest kernel
    to understand and implement.
Let ${Dx}$, ${Dy}$, and ${Dz}$ denote the block dimensions
    in the ${X}$, ${Y}$, and ${Z}$ axes, respectively.
Thus the block size is ${Dx \times Dy \times Dz}$.
Because we launch each kernel with a thread block of the same size,
    the total number of points must be $\leq$ ${1024}$ to respect the GPU limit of at most ${1024}$ threads per block. 
The GPU grid size of each data region is
    ${\lceil{Nx/Dx}\rceil \times \lceil{Ny/Dy}\rceil \times \lceil{Nz/Dz}\rceil}$.

During execution of the $25$-point stencil kernel,
    each thread fetches the point for itself,
    as well as $4$ neighboring points along each direction of each axis.
For good performance,
    we ensure a good memory access pattern 
    when stencil points are fetched directly from global memory.
Since we store the 3D grid data as a flat 1D array,
    we ensure global memory coalescing
    for the most innermost dimension ${X}$.

We refer to the family of 3D kernel implementations that fetch stencil points directly from the $u$ array in global memory as
    as \texttt{gmem\_\{Dx\}\_\{Dy\}\_\{Dz\}} in our experiments.

\subsubsection{3D Blocking Using Shared Memory for the $u$ Array}
This approach is a variant of the aforementioned 3D blocking using global memory. It uses same 3D blocking strategy for each of the data regions; however, instead of computing directly on data fetched from global memory,
 this implementation fetches the 
    ${u}$ array from global memory,
    stores it into shared memory,
    and performs the stencil computation on data fetched from shared memory.
The total number of points we fetch in this case is
    ${Dx \times Dy \times Dz}$ for a block and
    ${(Dx \times Dy + Dx \times Dz + Dy \times Dz) \times R \times 2}$
        for halos around the block.
For high order stencils, one must account for the halo size to ensure both the block and the halo fit in shared memory.

For high-order stencils,
    the halo accounts for a significant fraction of the data to fetch into shared memory.
Thus, designing the right approach to minimize the fetch cost
    is crucial to overall performance.
We describe the most general approach with good performance based on our experiments.
    
First, thread ${(i, j, k)}$ fetches the point ${(i, j, k)}$.
Then, when we design the block size,
    for each thread dimension,
    we must have at least ${2R}$ threads,
    and we use the first ${2R}$ threads along each dimension to fetch the halos.
Along each dimension, threads ${0}$ to ${R - 1}$ fetch the halo on one side,
    and threads ${R}$ to ${2R - 1}$ fetch the halo on the other side.
Fetching is perfectly balanced for each thread when
    ${ |D| = 2R }$.
    
To use this strategy for the acoustic isotropic model where ${R}$ is ${4}$,
    we need to have at least eight threads along each dimension.
Considering that maximum number of threads in a block is 1024
    and the total amount of shared memory per block,
    the only possible tile size for this case
        is $8 \times 8 \times 8$.
This results in perfectly balanced fetch work and computation for threads along each dimension.

Both the point and halo fetching need to be done in a fashion that enables global memory coalescing  to the greatest extent possible.

We refer to the implementation that uses 3D blocking and shared memory as ${smem\_u}$ in our experiments.

\subsubsection{3D Blocking Using Shared Memory for Boundary Regions}
This implementation also
 exploits shared memory
    and uses 3D blocking.
The main difference between this approach and the previous one is that
    this implementation fetches the $eta$ array into shared memory,
    whereas the previous approach fetches the $u$ array into shared memory.
Because $eta$ is only used
    in the stencil computation inside the PML region,
    this strategy applies only  in the PML kernel.

This approach may appear to be nothing new; however, it is interesting for two reasons.
First, as previously described,
     computations on $eta$  in the PML region use a low-order 7-point stencil
        rather than the 25-point high-order stencil of the inner region.
In fact, the halo size of $eta$ is just one. Such low-order stencils have been widely studied in the literature; however, the combination of high and low order stencils is seldom addressed.
Second, this implementation gives us an opportunity to observe
    the performance changes 
    by using global memory with a good access pattern
        for a high-order stencil,
    meanwhile using shared memory for a lower-order stencil.

In terms of fetching $eta$ into shared memory,
    we have two implementations that differ in the number of conditionals.
In our experiments, we refer to the shared memory kernel implementation that uses three conditionals as ${smem\_eta\_3}$ and the implementation that uses one conditional as
 ${smem\_eta\_1}$.
We let ${R\_eta}$ denote the width of halos for $eta$,
    and for the acoustic isotropic PML layer, $R\_eta$ is $1$.

${smem\_eta\_3}$ uses an approach similar to ${smem\_u}$,
    where the first ${2R\_eta}$ threads from each dimension
        fetch the halos.
Since we have three dimensions,
    we need three conditionals,
    one for each dimension, respectively.
Because ${R\_eta}$ is just ${1}$,
    we only need two threads fetching halos along each thread dimension.
This could introduce unbalanced fetch work
    because for a 3D block of \texttt{8x8x8},
    if only two threads in each dimension perform halo fetching,
    that is one fourth of the threads.
Because we have three dimensions,     
    only ${1/64}$ of the threads perform fetching,
    while others are idle.
    
\begin{algorithm}
	\KwData{$\mathbf{xidz, yidx, zidx}$: thread index of x, y, and z dimension, respectively}
	\KwData{$\mathbf{nt}$: number of threads per block dimension}
	\KwResult{$\mathbf{g}$: coordinate for global memory}
	\KwResult{$\mathbf{s}$: coordinate for shared memory}

	\If{zidx $ < 6$}{
	    z $\leftarrow$ zidx $\&$ $1$\;
	    sz $\leftarrow z * 9$\;
	    gz $\leftarrow z * (bt+1) - 1$\;

	    xzswap $\leftarrow$ zidx $<= 1$\;
	    yzswap $\leftarrow$ (zidx $\&$ $2$) $== 2$\;

	    si $\leftarrow$ xzswap $?$ sz $: ($xidx$+1)$\;
	    sj $\leftarrow$ yzswap $?$ sz $: ($yidx$+1)$\;
	    si $\leftarrow$ xzswap $? ($xidx$+1): ($yzswap$?($yidx$+1) : $ sz $)$\;

	    gi $\leftarrow$ xzswap $?$ gz $: $ xidx\;
	    gj $\leftarrow$ yzswap $?$ gz $: $ yidx\;
	    gi $\leftarrow$ xzswap $?$ xidx $: ($yzswap $?$ yidx$ : $ gz $)$\;

	    $\mathbf{s} \leftarrow $ (si, sj, sk)\;
	    $\mathbf{g} \leftarrow $ (gi, gj, gk)\;
    }
	\caption{Shared Memory Fetching Strategy for $eta$-array Using Only One Conditional}
	\label{algo:eta-1-fetching}
\end{algorithm}

To address the work imbalance, we propose ${smem\_eta\_1}$ with only one condition,
    where we choose to use the first six threads from the \texttt{X} dimension
        to perform halo fetching.
Algorithm \ref{algo:eta-1-fetching} describes how we tilt the six planes of threads
    to identify the halo point that the each thread is responsible for the fetching.
For $8 \times 8 \times 8$ 3D blocks,
    because ${6/8}$ threads perform the fetching,
    in theory, we reduce thread idleness to just ${20\%}$.
However, this algorithm has relatively complex arithmetic to tilt each thread to its proper halo position, so
an experimental evaluation is needed to see whether the strategy is profitable.

\subsubsection{Semi-stencil}

Semi-stencil was initially introduced for CPUs,
    where it separates the stencil computations into two phases,
    namely the forward phase and the backward phase.
The algorithm reads ${R+1}$ points on one dimension,
    and the forward phases compute as the points are the left side of the stencil,
    and the partial result is stored to the rightmost point.
Backward phases then compute as if the points are the right side of the stencil,
    and write out the final result to the leftmost point.
By doing this, semi-stencil has the potential
    to improve performance by changing the ratio between load and store.
For example, for a 3D stencil of halo size of ${R}$,
    with typical approach,
    it requires to load ${6 * R + 1}$ points
        in order to perform stencil computation for one point.
Once the computation is done, a single store writes the result back.
So the load-store-ratio is ${(6*R+1) : 1}$.
On the other hand, 
    using semi-stencil on one dimension,
    we read the center point and the half points with ${R+1}$ loads,
    then forward phase writes one store,
    and backward phase write another store,
    thus total of two stores.
Therefore, the load-store-ratio for semi-stencil is $(R+1):2$.
Also please note that, even for multi-dimensions,
    this ratio does not change.
This is in theory very appealing for high-order stencils
    because the larger the size of the halo,
    the potential benefits one might achieve
    as the algorithm trades half of loads with just one more store.
Therefore, we try to adopt semi-stencil algorithm on GPUs.

Our GPU implementation also uses 3D blocking.
While the stencil computation in each 3D block
    is close to the CPU implementations described in the original paper.
    we further parallelize the executions by
        running all time steps concurrently on GPUs.

We assign the identifier ${semi}$ to this implementation.

\subsubsection{2.5D Streaming with Multi-Plane using Shared Memory}

Starting from this implementation,
    we use 2.5D blocking.
As already described,
    the 2.5D algorithm mainly streams a 2D plane through the third dimension.
In our implementations,
    we choose the 2D XY-subplane
    because X is the innermost dimension in our data layout.
Let ${Dx}$ and ${Dy}$ denote the dimensions of the 2D tile along
    the $X$ and $Y$ axes, respectively.
So we launch kernels using 2D thread blocks with the size of $Dx$ by $Dy$ and a total of ${Dx \times Dy}$ threads.
The GPU grid size of each data region is
    $\lceil{Nx / Dx}\rceil \times \lceil{Ny / Dy}\rceil$.

In this approach, we exploit shared memory as a buffer
    to store all data needed in the stencil computations
    for a particular XY-subplane.
In addition to the current XY-subplane,
    we also load $R$ subplanes above the current subplane
    and $R$ subplanes below into the shared memory.
Therefore, we allocate a buffer for ${2R+1}$ planes,
    where each has ${ (Dx+2R) \times (Dy+2R) }$ points,
    thus, total of ${ (2R+1) \times (Dx+2R) \times (Dy+2R) }$ points.
Hence, the extent of each subplane must be carefully chosen so
 that the buffer size is as large as possible to enhance data reuse,
    but at the same time,
    it must be chosen so that the aggregate data volume of the planes doesn't exceed the shared memory available to a block.
Let $B$ denote our buffer, and $B[i]$ denote the i-th subplane in the buffer.

Before we can start the streaming computation,
    points from the top halos are pre-loaded into buffer ${B[0 .. R)}$ and the first $R$ XY-subplanes are pre-loaded into ${B[R .. 2R)}$.
Then, in our streaming loop, for each ${z \leftarrow [0 .. Nz)}$,
    we first load the \texttt{(z+R)}-th XY-subplane
        into ${B[(z+R)\mod(2R+1)] }$;
    next, we perform the stencil computation
        for the \texttt{z}-th XY-subplane
        with the stencil points read from $B$ in shared memory;
     finally, we store the result back to global memory.

While we explain this strategy using a modulus operator, in practice,
we avoid using it for speed.
Since the \texttt{z} index always increases by one    
    inside the streaming loop,
we use loop unrolling and index rotation
to achieve the desired effect without modulus computations.

We refer to the family of kernel implementations of this strategy as
    $st\_smem\_\{Dx\}\_\{Dy\}$ in our experiments.

The shared memory buffer used by this approach is limited by the GPU shared memory size.
Alternatively, one can store data for  stencil points along the streaming dimension in registers.
We discuss two approaches that use  registers to store points along the streaming dimension in the following sections.

\subsubsection{2.5D Streaming using Register Shifting}

In this 2.5D streaming approach, we keep
     points of the current XY-subplane in shared memory.
However, when we stream along the z-axis, we use registers for the  points along the z-axis.
In contrast to shared memory,
    where data loaded from one thread
        is accessible by other threads in the same block,
    registers are only accessible by the current thread.
Since we are streaming along the z-axis,
    the data from z-axis loaded for one thread
        is not needed by other threads.

So we allocate a shared memory space
    to hold ${ (Dx+2R) \times (Dy+2R) }$ points
    for the currently active plane.
The shared memory footprint compared to the previous method is
    ${ 1 : (2R + 1) }$.
For high-order stencils, $R$ is large so that the shared memory usage reduction is significant.
Let $S(x,y)$ denotes the shared memory with location $(x,y)$.

We also allocate $2R+1$ registers for the current point
    and its neighbors in each direction along the z-axis.
Let $Reg(x,y)[i]$ denote the \texttt{i}-th register
    for the thread $(x,y)$.

Before we can start the streaming computation,
    thread $(x,y)$ fetches data values from $(x,y,z)$ for ${z \leftarrow [-R .. R)}$
        and stores them into register $Reg(x,y)[0..2R)$, respectively.
Then, inside the stream loop, for each ${z \leftarrow [0 .. Nz)}$,
    we first shift the register indices back one position on each thread,
        such that for ${r \leftarrow (0 .. 2R]}$,
        ${ Reg(x,y)[r-1] = Reg(x,y)[r] }$.
    Then, we load the leading point
        along the streaming dimension $(x,y,z+R)$
        into register ${ Reg(x,y)[2R] }$. Next,
  we fetch data $(x,y,z)$ from global memory
        into $S(x,y)$;
    and we finally perform the stencil computation by
        using the data of XY-subplane from shared memory
        and data along the z-axis from registers.
    Finally, the kernel stores the stencil  result for each thread back to global memory.

We refer to the family of kernel implementations using this strategy as
    $st\_reg\_shft\_\{Dx\}\_\{Dy\}$ in our experiments.

Although the notation we use above for registers might give the impression that we are using array indexing to access register values,
in our implementation,
    registers are expressed explicitly as ${2R+1}$ scalar variables.
Since our acoustic isotropic kernel has $R=4$,
    which is the same as the sample code by Micikevicius~\cite{micike},
     our implementation uses the same variable names:
        \texttt{behind4},
        \texttt{behind3},
        \texttt{behind2},
        \texttt{behind1},
        \texttt{current},
        \texttt{front1},
        \texttt{front2},
        \texttt{front3},
        and \texttt{front4}.

\subsubsection{2.5D Streaming using Fixed Registers with Loop Unrolling}

Like the previous approach, this implementation uses
     shared memory for the current XY-subplane and registers for points along the z-axis -- the streaming dimension.
However, the values in the registers are fixed instead of being ``shifted.''

We again allocate a shared memory of ${ (Dx+2R) \times (Dy+2R) }$ points.
Let $S(x,y)$ denotes the shared memory for location $(x,y)$.
We allocate $2R+1$ registers as well,
    and denote $Reg(x,y)[i]$ for the \texttt{i}-th register
        of the thread $(x,y)$.
In practice, they are ${2R+1}$ named variables.

Before we can start the streaming computation,
    thread $(x,y)$ fetches data from $(x,y,z)$ for ${z \leftarrow [-R .. R)}$
        and stores them into register $Reg(x,y)[0..2R)$, respectively.
Then, inside the stream loop, for each ${z \leftarrow [0 .. Nz)}$,
    we do not modify any value in the existing registers; we only 
        update register $Reg(x,y)[(z+2R)\mod (2R+1)]$
        with the value of point $(x,y,z+R)$.
    Then, we fetch data $(x,y,z)$ from global memory
        into $S(x,y)$.
    Next, we perform the stencil computation by
        using the data of XY-subplane from shared memory
            and \texttt{i}-th data above current point
                from ${ Reg(x,y)[(z+R-i) \mod (2R+1)] }$,
            and \texttt{j}-th data below the current point
                from ${ Reg(x,y)[(z+R+j) \mod (2R+1)] }$. 
    Finally, the kernel stores the result back to global memory.

To further improve performance, we unroll the streaming loop.
We introduce macros
    with register indices as macro placeholders.
Inside the streaming loop,
    we expand $2R+1$ macro calls,
    each with register indices shifted by one.
We check and exit the loop
    when the stream reaches the boundary of z-axis.

We refer to the family of kernel implementations using this strategy  as
    $st\_reg\_fixed\_\{Dx\}\_\{Dy\}$ in our experiments.

\section{Evaluation}
\label{sec:evaluation}

\subsection{Experiment Environments}

We evaluate all kernel implementations
    on three machines with NVIDIA GPUs across several generations.
Table \ref{table:machine-spec} lists our machine specifications.
And we refer to the three machines by their GPU models.

\begin{table}[h!]
\centering
    \begin{tabular}{|c >{\centering\arraybackslash}m{1.5cm} >{\centering\arraybackslash}m{1.5cm} >{\centering\arraybackslash}m{1.5cm}|}
    \hline
     & V100 & P100 & NVS510 \\
    \hline
    CPU & IBM POWER9 & IBM POWER8NVL & Intel Xeon E3-1245 v6 \\
    \hline
    CPU Cores & 160 & 160 & 8 \\
    \hline
    RAM & 256 GB & 256 GB & 16 GB \\
    \hline
    GPU & NVIDIA Tesla V100 & NVIDIA Tesla P100 & NVIDIA NVS 510 \\
    \hline
    GRAM & 32 GB & 16 GB & 2 GB \\
    \hline
    OS & RHEL v7.7 & RHEL v7.4 & Ubuntu 18.04 LTS \\
    \hline
    CUDA & 10.2.89 & 10.1 & 10.2.89 \\
    \hline
    NVIDIA Driver & 440.33.01 & 418.39 & 440.33.01 \\
    \hline
    \end{tabular}
    \caption{Machine Specifications}
    \label{table:machine-spec}
\end{table}

\begin{itemize}
    \item 
Machine \texttt{V100} is equipped with four NVIDIA V100 GPUs. We
 use one dedicated GPU for our experiments.
We  use the compiler option \texttt{-arch=sm\_70} to compile all kernels for this platform. 
\item 
Machine \texttt{P100} is equipped with four NVIDIA P100 GPUs. We 
 use one dedicated GPU for our experiments.
We use the
    compiler option \texttt{-arch=sm\_60} to compile all kernels for this platform. 
\item Machine \texttt{NVS510} has one NVIDIA NVS510 GPU. 
We  use the compiler option \texttt{-arch=sm\_30} to compile all kernels for this platform. 
\end{itemize}

On \texttt{NVS510}, support
for some tooling is  marked
    as deprecated. While the tools work to some extent,
    many have limited functionality.
    Also,
 the GPU memory available on \texttt{NVS510} doesn't support
 grid sizes needed for real-world use.
Therefore, we only use this machine for basic comparisons
    across GPU generations.
While we examine some metrics on this platform, we don't discuss them in detail.

In most situations,
    we let the \texttt{nvcc} compiler
         figure out the register usage by itself,
    but we pay very close attention to the resulting register footprint.
However, there are a few cases,
    where we specify the maximum number of registers used by a kernel 
   using the compiler flag \texttt{-maxrregcount=X}
    to prevent register spilling.

We also use
    HPCToolkit \cite{hpctoolkit}, \cite{hpctoolkit-gpu} version 20200803
    and Empirical Roofline Toolkit \cite{roofline-ert}
    \footnote{We use a local fork of the tool
        for better Python 3 support and
        other minor changes needed for our environments.
        Our changes are made available
            at https://github.com/rsrice/cs-roofline-toolkit-fork
        under the same open-source license as the upstream repository
            https://bitbucket.org/berkeleylab/cs-roofline-toolkit/src/master/.},
    and NVIDIA Nsight Compute version 2019.5.0
    during our evaluations.

\subsection{Evaluation Methodologies}

We evaluate all implementations and their variants.
First, we conduct basic time measurements.
Second, we use HPCToolkit's GPU support~\cite{hpctoolkit-gpu} to profile the kernel details with PC sampling.
Third, we run Nsight Compute
    for device-specific kernel characteristics.
Finally, we use the Empirical Roofline Toolkit to understand memory bandwidth limits on algorithm performance.
We calculate the arithmetic intensity
    and the performance of every kernel,
    and compare them with the roofline chart.
We describe each of our evaluation methods below.

\subsubsection{Time Measurements}

For each machine, based on its device memory size,
    we run the kernels with a large grid size
       supported by the device memory.
For \texttt{V100}, we use a grid size of ${1000^3 }$;
    for \texttt{P100}, we use  a grid size of $893^3$;
    for \texttt{NVS510}, we use a grid size of $300^3$.
As described in \ref{sec:seismic-modeling-acoustic-iso},
    the stencil needs multiple iterations to converge. For benchmarking, we use $1000$ iterations
    for all kernels on all machines.
For each execution,
    we warm up the kernel by running the entire execution once,
    and then we repeat it five times, recording
 the average time for the five runs.

\subsubsection{HPCToolkit}

We use the August 2020 release of the HPCToolkit
    to collect GPU kernel metrics,
        such as register use, block and grid size,
    as well as PC sampling statistics
        such as exposed latencies and their kinds.

The evaluation contains four steps:
Firstly, running \texttt{hpcrun} along with kernel executions
    for sampling using program counters.
Thanks to the very low overhead of \texttt{hpcrun},
    we can run our kernels with 1000 time iterations,
    which gives us accurate measurements that
        match the kernel behavior in the real world.
Then, we use \texttt{hpcstruct} on the kernel binaries
    and recover the information
        about their relations to the source code.
This is needed to contribute the performance metrics
    back to the source code,
    so that we can evaluate it later at the source code level,
    which makes the investigation easier.
The source code structure computed by \texttt{hpcstruct}
    is then associated by \texttt{hpcprof}
    with the raw sampling data from \texttt{hpcrun}.
Finally, a HPCToolkit performance database is generated.

HPCToolkit provides two graphicl user interfacea 
    to analyze the performance database,
    namely, \texttt{HPCViewer} and \texttt{HPCTraceViewer}.
We use \texttt{HPCViewer} primarily for
    issues such as memory stalls, which enables us to easily
  spot which source lines have the most significant stalls.
We also use \texttt{HPCViewer} to quickly identify
    the performance hotspots in our kernel executions using its
 code-centric views.
We use \texttt{HPCTraceViewer}
    to inspect the program execution over time, which enables us to quickly spot idleness and see the associated calling contexts. 
    
We  describe some of our findings using HPCToolkit
    in our discussion of the evaluation results.

\subsubsection{Nsight Compute}

We run Nsight Compute for kernel characteristics,
    such as theoretical and achieved occupancy.
Nsight Compute provides insights
    when performance differences
    are driven by the kernel characteristics.
For example, when low occupancy happens,
    one can easily tell from an Nsight Compute report 
        whether or not the problem seems to be associated with 
        the register footprint, the shared memory footprint,
    or the number of threads.

Nsight Compute re-plays every kernel execution multiple times
to collect a complete set of measurements, which adds a huge measurement overhead. When we use Nsight Compute, we run only five iterations.

\begin{table*}[h]
    \begin{center}
    \rowcolors{2}{gray!25}{white}
\begin{tabular}{ |c||c|c|c|c||r|r|r|  }
\hline
\multicolumn{5}{|c||}{Kernel} & \multicolumn{3}{|c|}{Machine} \\
\hline
Kernel Identifier & Dx & Dy & Dz & Nr & V100 & P100 & NVS510 \\
\hline
gmem\_4x4x4              & 4  & 4  & 4 & -  & 77.77 & 181.99 & 682.89 \\
gmem\_8x8x4              & 8  & 8  & 4 & -  & 71.91 & 167.75 & 674.09 \\
gmem\_8x8x8              & 8  & 8  & 8 & -  & 53.88 & 117.74 & 415.85 \\
gmem\_16x16x4            & 16 & 16 & 4 & -  & 85.52 & 195.82 & 760.72 \\
gmem\_32x32x1            & 32 & 32 & 1 & -  & 292.36  & 639.62  & 2507.22 \\
smem\_u                  & 8  & 8  & 8 & -  & 57.30 & 76.18 & 210.42 \\
smem\_eta\_1             & 8  & 8  & 8 & -  & 54.87 & 119.15 & 397.56 \\
smem\_eta\_3             & 8  & 8  & 8 & -  & 54.34 & 117.39 & 396.49 \\
semi                     & 8  & 8  & 8 & -  & 172.84 & 217.29 & 1726.17 \\
st\_smem\_8x8           & 8  & 8  & - & -  & 116.38  & 112.71 & 509.18 \\
st\_smem\_8x16          & 8  & 16 & - & -  & 113.46  & 105.41  & 439.47 \\
st\_smem\_16x8          & 16 & 8  & - & -  & 59.92 & 77.91 & 425.73 \\
st\_smem\_16x16         & 16 & 16 & - & -  & 55.87 & 72.73 & 349.45 \\
st\_reg\_shft\_8x8        & 8  & 8  & - & -  & 104.36 & 144.89  & 209.87 \\
st\_reg\_shft\_16x16      & 16 & 16 & - & -  & 65.79 & 80.23 & 182.52 \\
st\_reg\_shft\_16x32      & 16 & 32 & - & -  & 65.61 & 82.25 & 199.61  \\
st\_reg\_shft\_16x64      & 16 & 64 & - & 64 & 115.54 & 98.19 & 240.41 \\
st\_reg\_shft\_32x16      & 32 & 16 & - & -  & 60.83 & 70.63 & 171.30 \\
st\_reg\_shft\_32x32      & 32 & 32 & - & 64 & 93.92  & 76.27 & 167.29 \\
st\_reg\_shft\_64x16      & 64 & 16 & - & 64 & 90.98  & 80.67 & 202.74 \\
st\_reg\_fixed\_8x8            & 8  & 8  & - & -  & 113.88 & 152.75 & 195.05 \\
st\_reg\_fixed\_16x8           & 16 & 8  & - & -  & 70.24 & 84.05  & 159.73 \\
st\_reg\_fixed\_16x16          & 16 & 16 & - & -  & 61.66 & 76.10 & 170.03 \\
st\_reg\_fixed\_32x16          & 32 & 16 & - & -  & 62.45 & 66.60 & 162.05 \\
st\_reg\_fixed\_32x32          & 32 & 32 & - & 64 & 58.96   & 61.74   & 160.91 \\
\hline
\end{tabular}
\caption{Time Measurement}
\label{table:time-measurement}
\end{center}
\end{table*}

\subsubsection{Roofline Performance Model}

We use the GPU Roofline performance model to see how well our kernels perform relative to a machine's practical peak based on each kernel's arithmetic intensity and the memory  bandwidth-based performance limit for that particular arithmetic intensity.

We use the Empirical Roofline Toolkit (ERT) for machine characterizations.
It runs several micro-benchmarks
    to characterize the peak compute speed and memory bandwidth
    of the machine.
Benchmarking directly on a machine gives us an achievable performance bound, which is substantially lower than the theoretical peak claimed by the manufacturers when a kernel is memory bound.

We characterize kernels using \texttt{nvprof}  by measuring
    several kernel performance metrics,
including FLOPs,
        L2 read and write transactions,
        as well as DRAM read and write transactions.
Output from \texttt{nvprof} is then fed into the calculations of
    both the performance and the arithmetic intensity for each kernel.
Performance is calculated by the division of the measured FLOPS by the measured execution time.
Arithmetic intensities are calculated by the division of the measured FLOPS by the measured bytes accessed on DRAM and L2 cache respectively.

We compare the performance of each kernel with
    the peak performance of the machine it runs on.
We then compare kernels by their arithmetic intensities
    and their relative performance.

\subsection{Results}

In this section, we first present a summary of our results in tables and plots. After presenting our findings, we discuss our kernel measurements from several perspectives. 

Table \ref{table:time-measurement} presents time measurements for the kernels.
For 3D blockings, the columns $Dx$, $Dy$, and $Dz$
    stand for the block dimensions along the
     x, y, and z axes, respectively.
For 2.5D blockings, only columns for $Dx$ and $Dy$ are reported since the z-axis is unpartitioned.
For $Nr$ column,
    only values we explicitly specified are reported,
    and we use \texttt{-} for the ones that compiler decide.

\begin{table*}[h!]
    \begin{center}
    \rowcolors{2}{gray!25}{white}
\begin{tabular}{ |c||
    >{\raggedleft\arraybackslash}p{1.5cm}|
    >{\raggedleft\arraybackslash}p{1.5cm}|
    >{\raggedleft\arraybackslash}p{1.5cm}|
    >{\raggedleft\arraybackslash}p{1.5cm}|
    >{\raggedleft\arraybackslash}p{1.5cm}|
    >{\raggedleft\arraybackslash}p{1.5cm}|
    >{\raggedleft\arraybackslash}p{1.5cm}| }
\hline
Kernel Identifier   & 
\multicolumn{1}{>{\centering\arraybackslash}p{1.5cm}|}{Block Size} & 
\multicolumn{1}{>{\centering\arraybackslash}p{1.5cm}|}{Grid Size} & 
\multicolumn{1}{>{\centering\arraybackslash}p{1.5cm}|}{Registers Per Thread} & 
\multicolumn{1}{>{\centering\arraybackslash}p{1.5cm}|}{Achieved Active Warps} & 
\multicolumn{1}{>{\centering\arraybackslash}p{1.5cm}|}{Achieved Occupancy} & 
\multicolumn{1}{>{\centering\arraybackslash}p{1.5cm}|}{Theoretical Active Warps} & 
\multicolumn{1}{>{\centering\arraybackslash}p{1.5cm}|}{Theoretical Occupancy} \\
\hline
gmem\_4x4x4          & 64        & 13,312,053  & 40      & 37.2   & 58.2    & 48.0  & 75.0 \\
gmem\_8x8x4          & 256       & 3,356,157   & 40      & 44.0    & 68.7   & 48.0  & 75.0 \\
gmem\_8x8x8          & 512       & 1,685,159   & 40      & 42.5   & 66.4    & 48.0  & 75.0 \\
gmem\_16x16x4        & 1,024     & 853,200     & 40      & 28.9   & 45.2    & 32.0  & 50.0 \\
gmem\_32x32x1        & 1,024     & 851,400     & 40      & 29.3   & 45.8    & 32.0  & 50.0 \\
smem\_u              & 512       & 1,685,159   & 38      & 44.6   & 69.7    & 48.0  & 75.0 \\
smem\_eta\_1          & 512       & 1,685,159   & 40      & 42.4   & 66.3   & 48.0  & 75.0 \\
smem\_eta\_3          & 512       & 1,685,159   & 40      & 42.4   & 66.2   & 48.0  & 75.0 \\
semi                & 768       & 1,685,159   & 40      & 41.2   & 64.4     & 48.0  & 75.0 \\
st\_smem\_8x8       & 64        & 14,161      & 56      & 19.9   & 31.1     & 20.0  & 31.2 \\
st\_smem\_8x16      & 128       & 7,140       & 56      & 27.9   & 43.6     & 28.0  & 43.7 \\
st\_smem\_16x8      & 128       & 7,140       & 56      & 27.9    & 43.5    & 28.0  & 43.7 \\
st\_smem\_16x16     & 256       & 3,600       & 56      & 31.6   & 49.4     & 32.0  & 50.0 \\
st\_reg\_shft\_8x8    & 64        & 14,161      & 96      & 19.0   & 29.7   & 20.0  & 31.2 \\
st\_reg\_shft\_16x16  & 256       & 3,600       & 96      & 15.9   & 24.9   & 16.0  & 25.0 \\
st\_reg\_shft\_16x32  & 512       & 1,800       & 96      & 16.0    & 25.0  & 16.0  & 25.0 \\
st\_reg\_shft\_16x64  & 1,024     & 900         & 64      & 32.0    & 50.0  & 32.0  & 50.0 \\
st\_reg\_shft\_32x16  & 512       & 1,800       & 96      & 16.0    & 25.0  & 16.0  & 25.0 \\
st\_reg\_shft\_32x32  & 1,024     & 900         & 64      & 32.0    & 50.0  & 32.0  & 50.0 \\
st\_reg\_shft\_64x16  & 1,024     & 900         & 64      & 32.0    & 50.0  & 32.0  & 50.0 \\
st\_ref\_fixed\_8x8    & 64       & 14,161      & 78      & 23.9   & 37.3   & 24.0  & 37.5 \\
st\_ref\_fixed\_16x8   & 128      & 7,140       & 78      & 23.9   & 37.3   & 24.0  & 37.5 \\
st\_ref\_fixed\_16x16  & 256      & 3,600       & 78      & 23.9   & 37.4   & 24.0  & 37.5 \\
st\_ref\_fixed\_32x16  & 512      & 1,800       & 78      & 16.0   & 25.0   & 16.0  & 25.0 \\
st\_ref\_fixed\_32x32  & 1,024    & 900         & 64      & 32.0   & 50.0   & 32.0  & 50.0 \\
\hline
\end{tabular}
\end{center}

\begin{center}
\rowcolors{2}{gray!25}{white}
\begin{tabular}{ |c||
    >{\raggedleft\arraybackslash}p{0.78cm}|
    >{\raggedleft\arraybackslash}p{0.78cm}|
    >{\raggedleft\arraybackslash}p{0.78cm}|
    >{\raggedleft\arraybackslash}p{0.78cm}|
    >{\raggedleft\arraybackslash}p{0.78cm}|
    >{\raggedleft\arraybackslash}p{0.78cm}|
    >{\raggedleft\arraybackslash}p{0.78cm}|
    >{\raggedleft\arraybackslash}p{0.78cm}|
    >{\raggedleft\arraybackslash}p{0.78cm}|
    >{\raggedleft\arraybackslash}p{0.78cm}|
    >{\raggedleft\arraybackslash}p{0.78cm}|
    >{\raggedleft\arraybackslash}p{0.78cm}|
    >{\raggedleft\arraybackslash}p{0.78cm}|
    }
\hline
 & \multicolumn{4}{c|}{Static} &
   \multicolumn{3}{c|}{Top/Bottom} &
   \multicolumn{3}{c|}{Front/Back} &
   \multicolumn{3}{c|}{Left/Right} \\
\hline
Kernel Identifier   &
Block Size &
\scriptsize{Registers Per Thread} &
\tiny{Theoretical Active Warps} &
\tiny{Theoretical Occupancy} &
Grid Size &
\scriptsize{Achieved Active Warps} &
\scriptsize{Achieved Occupancy} &
Grid Size &
\scriptsize{Achieved Active Warps} &
\scriptsize{Achieved Occupancy} &
Grid Size &
\scriptsize{Achieved Active Warps} &
\scriptsize{Achieved Occupancy} \\
\hline
gmem\_4x4x4            & 64      &  48   &  40.0 &  62.5   & 437500      &  38.0   & 59.5  & 414750  & 38.0  & 59.4  & 393183    &  38.2  &   59.7 \\
gmem\_8x8x4            & 256     & 48    &  40.0 &  62.5   & 109375      &  37.5   & 58.6  & 118500  & 37.5  & 58.6  & 112812    &  36.7  &   57.3 \\
gmem\_8x8x8            & 512     & 48    &  32.0 &  50.0   &  62500      &  29.2   & 45.7  & 59500   & 26.9  & 42.0  & 56644     &  28.2  &   44.1 \\
gmem\_16x16x4          & 1024    &    48 &  32.0 &  50.0   &  27783      &  30.0   & 46.6  & 29862   & 29.5  & 46.1  & 28440     &  26.0  &   40.2 \\
gmem\_32x32x1          & 1024    &    48 &  32.0 &  50.0   &  27648      &  29.0   & 45.0  & 30272   & 29.0  & 45.0  & 28380     &  24.2  &   38.0 \\
smem\_u                & 512     & 48    &  32.0 &  50.0   & 62500       &  30.1   & 47.1  & 59500   & 27.8  & 43.5  & 56644     &  27.7  &   43.3 \\
smem\_eta\_1           & 512     & 32    &  64.0 &  100.0  & 62500       &  59.4   & 92.9  & 59500   & 54.8  & 85.6  & 56644     &  54.8  &   85.7 \\
smem\_eta\_3           & 512     & 32    &  64.0 &  100.0  & 62500       &  59.1   & 92.4  & 59500   & 53.9  & 84.3  & 56644     &  54.2  &   84.8 \\
semi                   & 768     & 64    &  24.0 &  37.5   & 62500       &  17.7   & 27.6  & 59500   & 18.7  & 29.3  & 56644     &  17.5  &   27.3 \\
st\_smem\_8x8          & 64      &  72   &  20.0 &  31.2   & 500         &  12.4   & 19.4  & 476     & 11.8  & 18.5  & 14161     &  19.7  &   30.8 \\
st\_smem\_8x16         & 128     & 72    &  28.0 &  43.7   & 252         &  12.6   & 19.7  & 238     & 11.8  & 18.5  & 7140      &  27.5  &   43.1 \\
st\_smem\_16x8         & 128     & 72    &  28.0 &  43.7   & 250         &  12.4   & 19.5  & 240     & 11.9  & 18.6  & 7140      &  27.5  &   43.0 \\
st\_smem\_16x16        & 256     & 72    &  24.0 &  37.5   & 126         &  12.7   & 19.8  & 120     & 12.0  & 18.7  & 3600      &  23.9  &   37.3 \\
st\_reg\_shft\_8x8     & 64      &  80   &  24.0 &  37.5   & 500         &  12.4   & 19.4  & 476     & 11.8  & 18.4  & 14161     &  23.6  &   36.8 \\
st\_reg\_shft\_16x16   & 256     & 80    &  24.0 &  37.5   & 126         &  12.6   & 19.7  & 120     & 11.9  & 18.6  & 3600      &  23.9  &   37.3 \\
st\_reg\_shft\_16x32   & 512     & 80    &  16.0 &  25.0   &  64         &  16.0   & 25.0  & 60      & 16.0  & 25.0  & 1800      &  15.9  &   24.9 \\
st\_reg\_shft\_16x64   & 1024    &    64 &  32.0 &  50.0   &  32         &  32.0   & 50.0  & 60      & 31.9  & 49.9  & 900       &  31.9  &   49.8 \\
st\_reg\_shft\_32x16   & 512     & 80    &  16.0 &  25.0   &  63         &  16.0   & 25.0  & 60      & 16.0  & 25.0  & 1800      &  15.9  &   24.9 \\
st\_reg\_shft\_32x32   & 1024    &    64 &  32.0 &  50.0   &  32         &  32.0   & 50.0  & 30      & 31.9  & 49.9  & 900       &  31.8  &   49.8 \\
st\_reg\_shft\_64x16   & 1024    &    64 &  32.0 &  50.0   &  63         &  31.9   & 49.9  & 30      & 31.9  & 49.9  & 900       &  31.8  &   49.8 \\
st\_ref\_fixed\_8x8    & 64      &  106  &  16.0 &  25.0   &  500        &  12.4   & 19.4  & 476     & 11.8  & 18.4  & 14161     &  15.7  &   24.6 \\
st\_ref\_fixed\_16x8   & 128     & 104   &  16.0 &  25.0   &  250        &  12.4   & 19.5  & 240     & 11.8  & 18.5  & 7140      &  15.7  &   24.6  \\
st\_ref\_fixed\_16x16  & 256     & 104   &  16.0 &  25.0   &  126        &  12.6   & 19.8  & 120     & 12.0  & 18.7  & 3600      &  15.8  &   24.7 \\
st\_ref\_fixed\_32x16  & 512     & 106   &  16.0 &  25.0   &  63         &  16.0   & 25.0  & 60      & 16.0  & 25.0  & 1800      &  15.9  &   24.9 \\
st\_ref\_fixed\_32x32  & 1024    &    64 &  32.0 &  50.0   &  32         &  32.0   & 50.0  & 30      & 31.9  & 49.9  & 900       &  31.9  &   49.8  \\
\hline
\end{tabular}
\caption{Kernel Characteristics on V100: (top) Inner; (bottom) PML }
\label{table:kernel-characterstics-v100}
\end{center}
\end{table*}

Table \ref{table:kernel-characterstics-v100}
    present the kernel characteristics
    for inner data region at the top,
    and PML regions at the bottom, respectively.
As previous discussed, there are three symmetric groups for the PML subregions:
    top/bottom, front/back, and left/right.
We group them in Table \ref{table:kernel-characterstics-v100}.
For shared characteristics across data regions,
    we further extract them into the \texttt{Static} column.

\begin{table*}[h!]
    \begin{center}
    \rowcolors{2}{gray!25}{white}
\begin{tabular}{ |c||
    >{\raggedleft\arraybackslash}p{0.9cm}|
    >{\raggedleft\arraybackslash}p{1.1cm}||
    >{\raggedleft\arraybackslash}p{0.9cm}|
    >{\raggedleft\arraybackslash}p{1.0cm}|
    >{\raggedleft\arraybackslash}p{1.1cm}|
    >{\raggedleft\arraybackslash}p{1.0cm}||
    >{\raggedleft\arraybackslash}p{0.9cm}|
    >{\raggedleft\arraybackslash}p{1.0cm}|
    >{\raggedleft\arraybackslash}p{1.1cm}|
    >{\raggedleft\arraybackslash}p{0.9cm}|  }
\hline
Kernel Identifier &
\multicolumn{1}{>{\centering\arraybackslash}p{0.9cm}|}{FLOP (x$10^{13}$)} &
\multicolumn{1}{>{\centering\arraybackslash}p{1.1cm}||}{Achieved Performance (GFLOPs)} &
\multicolumn{1}{>{\centering\arraybackslash}p{0.9cm}|}{L2 Transactions (x$10^{12}$)} &
\multicolumn{1}{>{\centering\arraybackslash}p{1.0cm}|}{L2 Arithmetic Intensity} &
\multicolumn{1}{>{\centering\arraybackslash}p{1.1cm}|}{L2 Machine Peak Performance (GFLOPs)} &
\multicolumn{1}{>{\centering\arraybackslash}p{1.0cm}||}{L2 Achieved Percentage} &
\multicolumn{1}{>{\centering\arraybackslash}p{0.9cm}|}{DRAM Transactions (x$10^{11}$)} &
\multicolumn{1}{>{\centering\arraybackslash}p{1.0cm}|}{DRAM Arithmetic Intensity} &
\multicolumn{1}{>{\centering\arraybackslash}p{1.1cm}|}{DRAM Machine Peak Performance (GFLOPs)} &
\multicolumn{1}{>{\centering\arraybackslash}p{0.9cm}|}{DRAM Achieved Percentage} \\
\hline
gmem\_4x4x4\_opt           & 4.453    & 533  &   3.38 &  0.41    & 1361    & 39.19\%  &   8.42    & 1.65    & 1291    & 41.29\%  \\
gmem\_8x8x4\_opt           & 4.453    & 577  &   2.81 &  0.49    & 1635    & 35.27\%  &   7.26    & 1.92    & 1498    & 38.50\%  \\
gmem\_8x8x8\_opt           & 4.453    & 770  &   1.79 &  0.78    & 2566    & 30.00\%  &   7.26    & 1.92    & 1498    & 51.39\%  \\
gmem\_16x16x4\_opt         & 4.453    & 485  &   2.45 &  0.57    & 1877    & 25.83\%  &   6.67    & 2.08    & 1628    & 29.78\%  \\
gmem\_32x32x1\_opt         & 4.453    & 142  &  13.90 &  0.10    & 330     & 42.95\%  &   6.56    & 2.12    & 1656    & 8.57\%   \\
smem\_u\_opt               & 4.453    & 724  &   1.82 &  0.77    & 2531    & 28.60\%  &   7.37    & 1.89    & 1474    & 49.11\%  \\
smem\_eta\_1\_opt          & 4.453    & 756  &   1.82 &  0.76    & 2522    & 29.97\%  &   7.31    & 1.90    & 1487    & 50.81\%  \\
smem\_eta\_3\_opt          & 4.453    & 763  &   1.81 &  0.77    & 2535    & 30.10\%  &   7.31    & 1.90    & 1488    & 51.30\%  \\
semi\_opt                  & 6.400    & 345  &   2.67 &  0.75    & 2480    & 13.90\%  &  18.40    & 1.08    & 847     & 40.71\%  \\
st\_smem\_8x8\_opt         & 4.453    & 356  &   1.59 &  0.87    & 2891    & 12.33\%  &  12.30    & 1.13    & 885     & 40.27\%  \\
st\_smem\_8x16\_opt        & 4.453    & 366  &   1.47 &  0.95    & 3130    & 11.68\%  &  13.30    & 1.05    & 820     & 44.58\%  \\
st\_smem\_16x8\_opt        & 4.453    & 692  &   1.17 &  1.19    & 3933    & 17.59\%  &   7.74    & 1.80    & 1404    & 49.27\%  \\
st\_smem\_16x16\_opt       & 4.453    & 742  &   1.04 &  1.34    & 4414    & 16.81\%  &   6.97    & 2.00    & 1560    & 47.58\%  \\
st\_reg\_shft\_8x8\_opt    & 4.453    & 397  &   1.57 &  0.89    & 2935    & 13.54\%  &  10.40    & 1.34    & 1047    & 37.96\%  \\
st\_reg\_shft\_16x16\_opt  & 4.453    & 630  &   1.20 &  1.16    & 3841    & 16.41\%  &   7.22    & 1.93    & 1506    & 41.86\%  \\
st\_reg\_shft\_16x32\_opt  & 4.453    & 632  &   1.15 &  1.21    & 3991    & 15.84\%  &   6.76    & 2.06    & 1607    & 39.32\%  \\
st\_reg\_shft\_16x64\_opt  & 4.453    & 359  &   1.99 &  0.70    & 2317    & 15.49\%  &  17.00    & 0.82    & 638     & 56.25\%  \\
st\_reg\_shft\_32x16\_opt  & 4.453    & 682  &   0.94 &  1.47    & 4861    & 14.02\%  &   6.94    & 2.00    & 1566    & 43.54\%  \\
st\_reg\_shft\_32x32\_opt  & 4.453    & 442  &   1.67 &  0.83    & 2750    & 16.05\%  &  15.50    & 0.90    & 701     & 62.95\%  \\
st\_reg\_shft\_64x16\_opt  & 4.453    & 456  &   1.57 &  0.89    & 2938    & 15.52\%  &  14.50    & 0.96    & 752     & 60.64\%  \\
st\_reg\_fixed\_8x8\_opt   & 4.453    & 364  &   1.65 &  0.84    & 2791    & 13.05\%  &  15.00    & 0.93    & 723     & 50.36\%  \\
st\_reg\_fixed\_16x8\_opt  & 4.453    & 590  &   1.27 &  1.10    & 3632    & 16.26\%  &   9.59    & 1.45    & 1133    & 52.11\%  \\
st\_reg\_fixed\_16x16\_opt & 4.453    & 673  &   1.18 &  1.18    & 3899    & 17.25\%  &   7.71    & 1.80    & 1409    & 47.72\%  \\
st\_reg\_fixed\_32x16\_opt & 4.453    & 664  &   9.12 &  1.53    & 5043    & 13.17\%  &   7.14    & 1.95    & 1522    & 43.62\%  \\
st\_reg\_fixed\_32x32\_opt & 4.453    & 703  &   1.09 &  1.27    & 4209    & 16.71\%  &   9.08    & 1.53    & 1197    & 58.78\%  \\
\hline
\end{tabular}
\caption{Kernel Performance Characteristics on V100}
\label{table:performance-characterstics}
\end{center}
\end{table*}

\begin{figure*}[h!]
     \centering
     \begin{subfigure}[b]{0.49\textwidth}
         \centering
         \includegraphics[width=\textwidth]{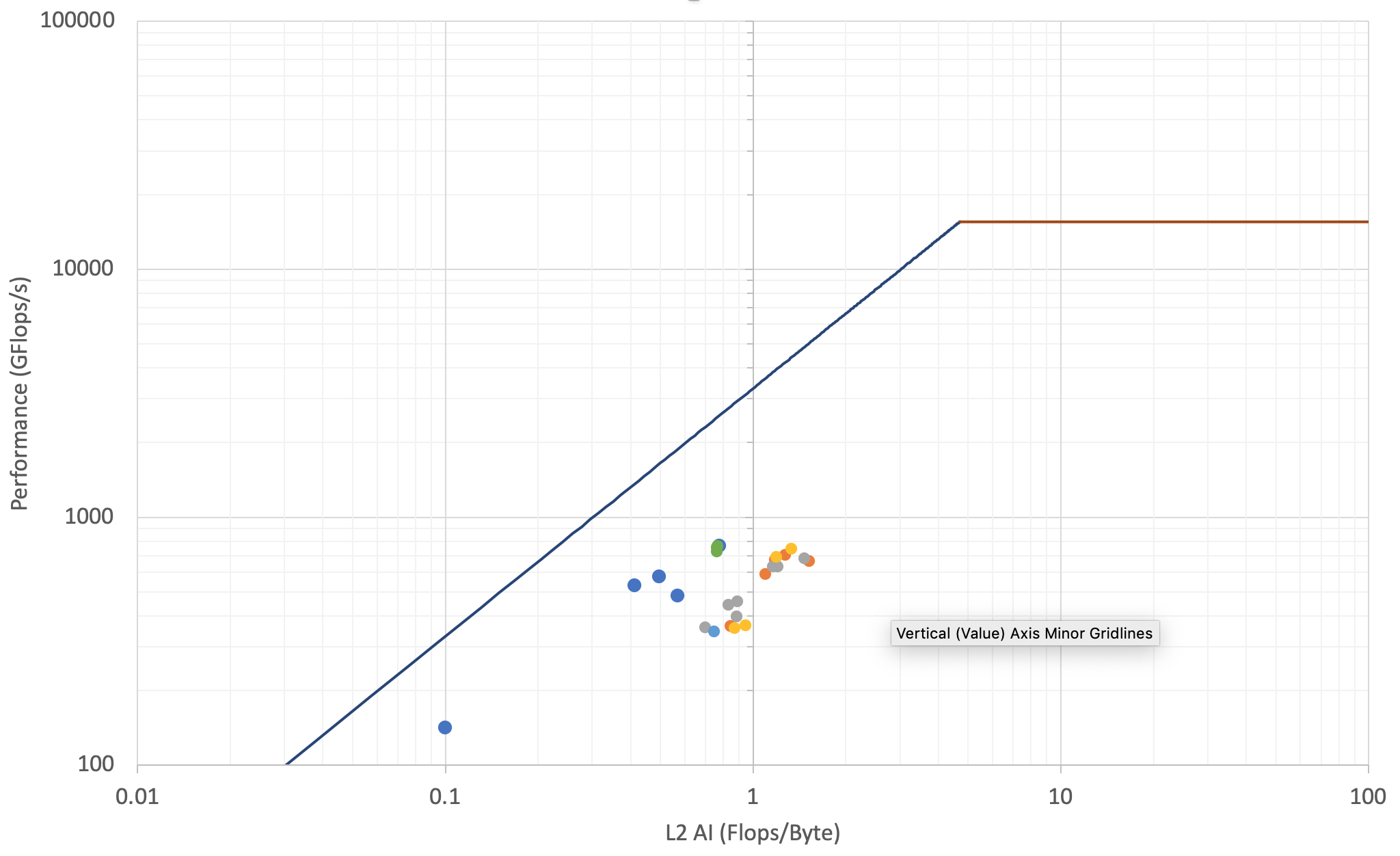}
         \caption{Device Performance vs L2 Arithmetic Intensity}
         \label{fig:roofline-device-l2}
     \end{subfigure}
     \hfill
     \begin{subfigure}[b]{0.49\textwidth}
         \centering
         \includegraphics[width=\textwidth]{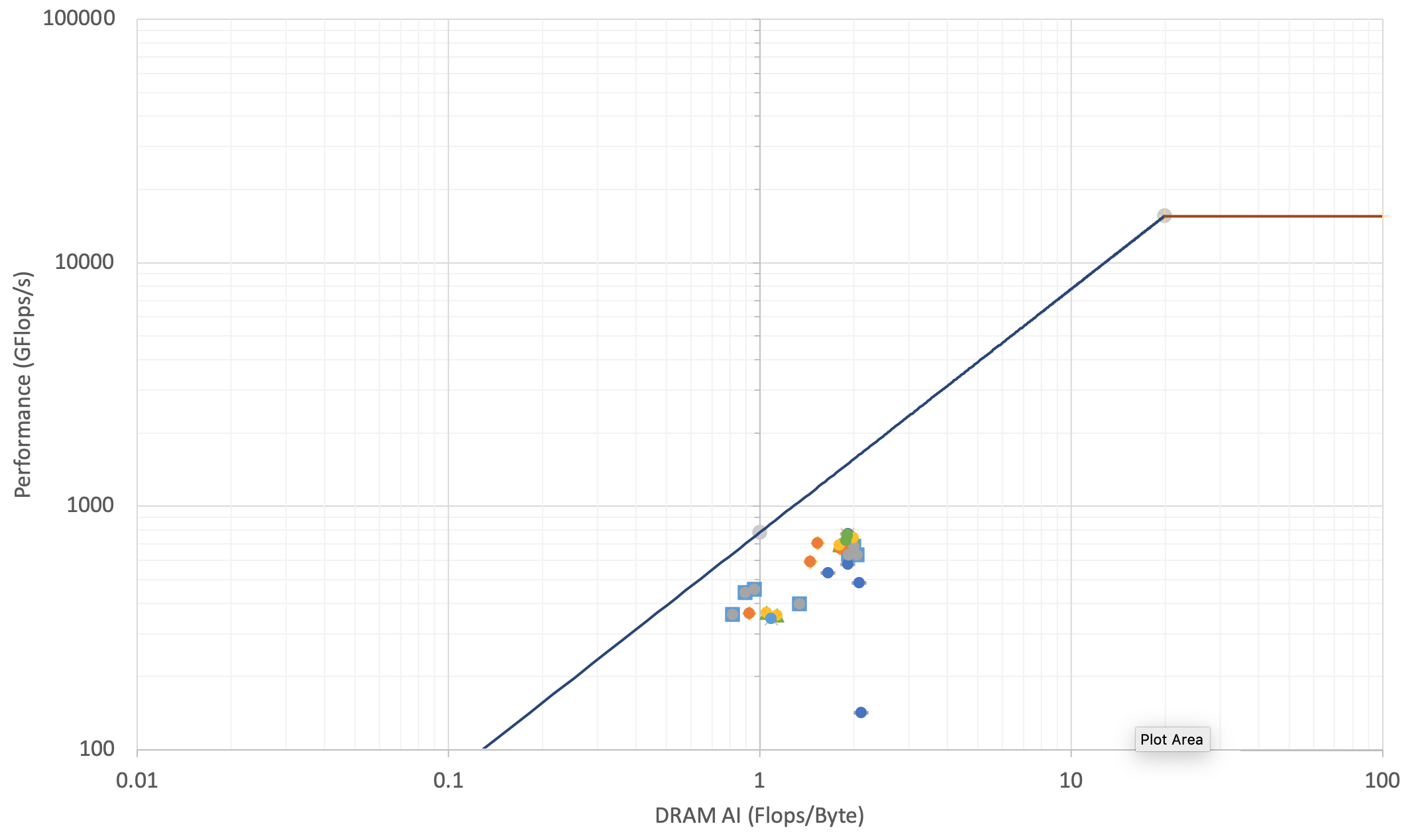}
         \caption{Device Performance vs DRAM Arithmetic Intensity}
         \label{fig:roofline-device-dram}
     \end{subfigure}
     \vfill
     \begin{subfigure}[b]{0.49\textwidth}
         \centering
         \includegraphics[width=\textwidth]{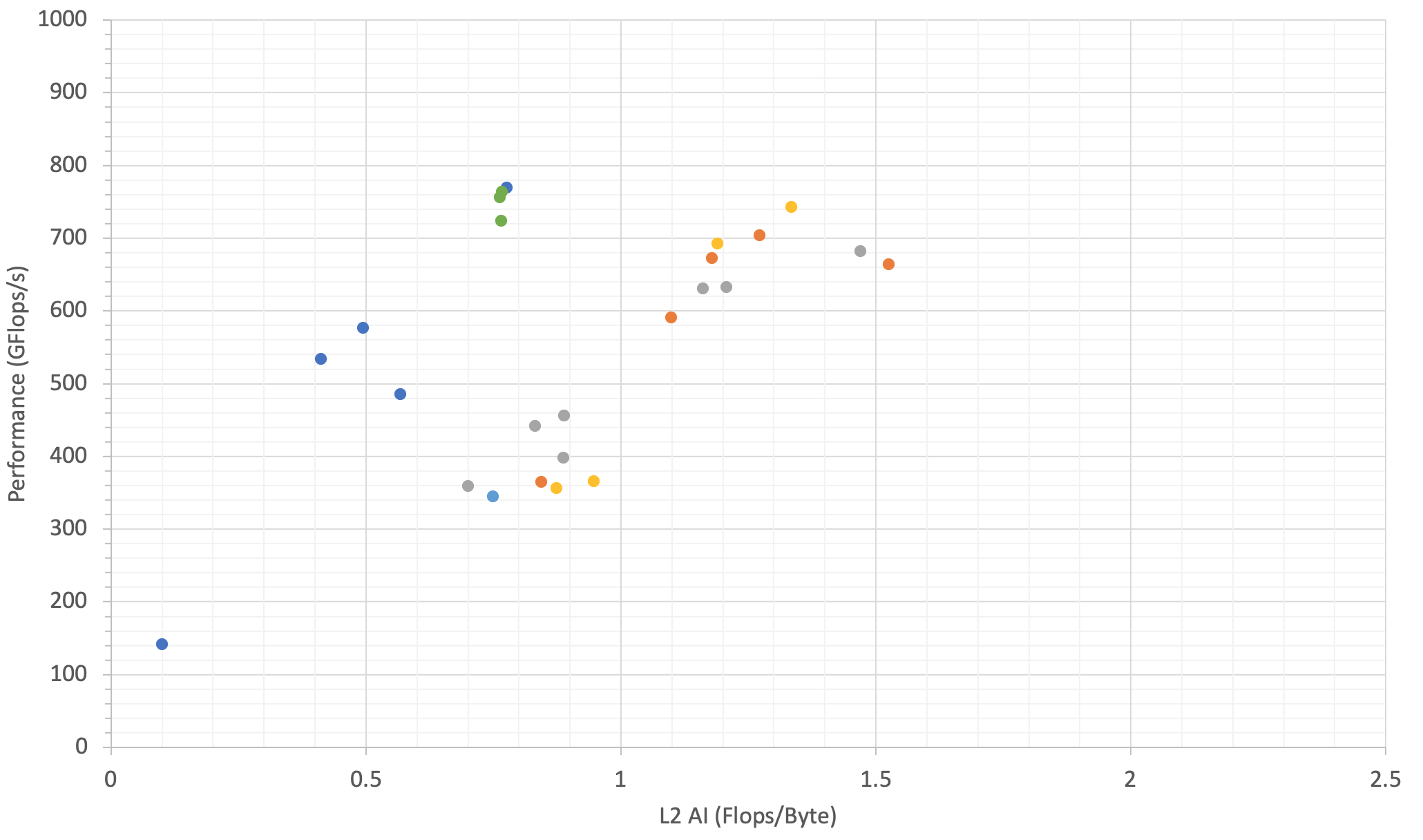}
         \caption{Kernel Performance vs L2 Arithmetic Intensity}
         \label{fig:roofline-kernel-l2}
     \end{subfigure}
     \hfill
     \begin{subfigure}[b]{0.49\textwidth}
         \centering
         \includegraphics[width=\textwidth]{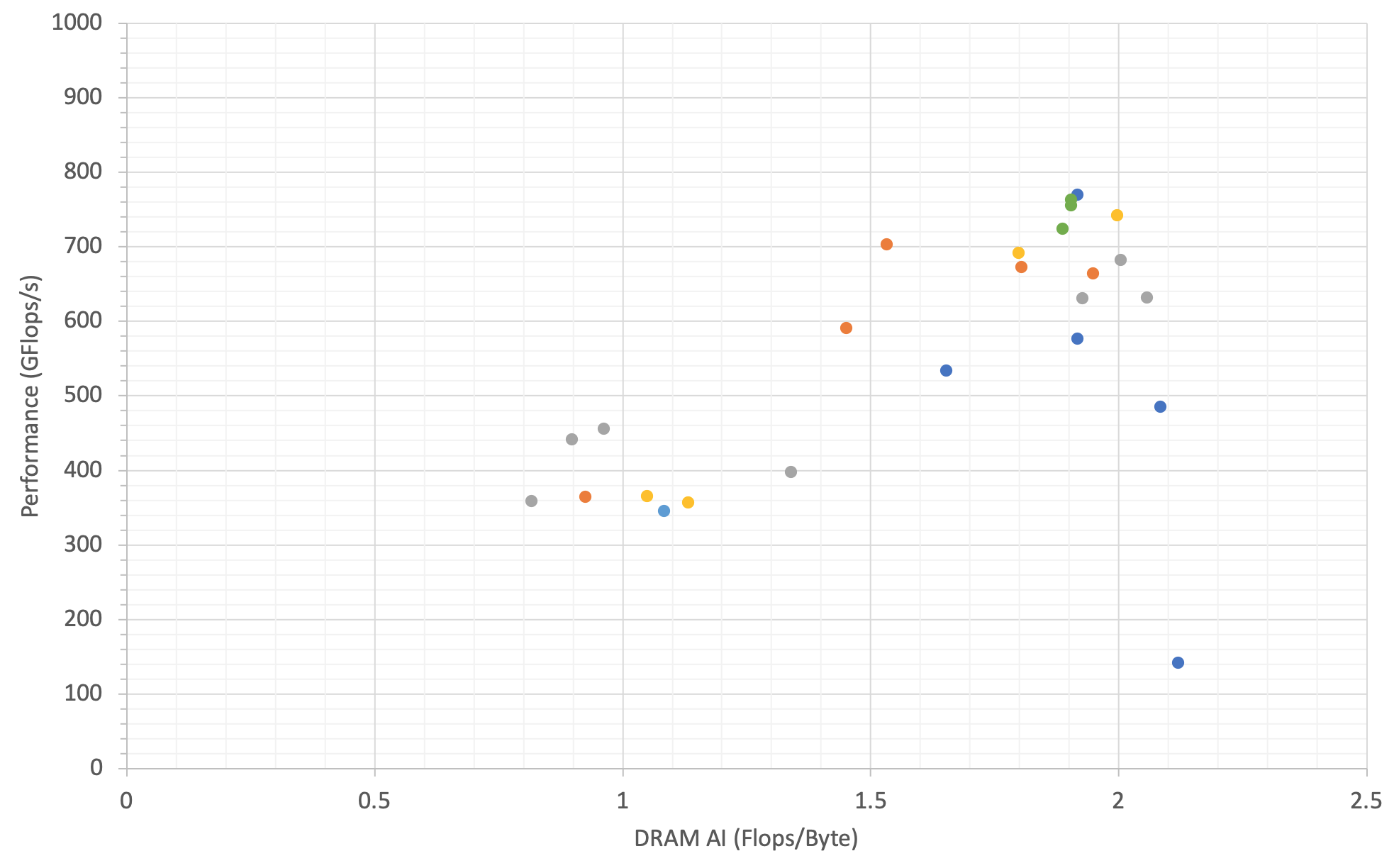}
         \caption{Kernel Performance vs DRAM Arithmetic Intensity}
         \label{fig:roofline-kernel-dram}
     \end{subfigure}
     \vfill
     \begin{subfigure}[b]{0.6\textwidth}
         \centering
         \includegraphics[width=\textwidth]{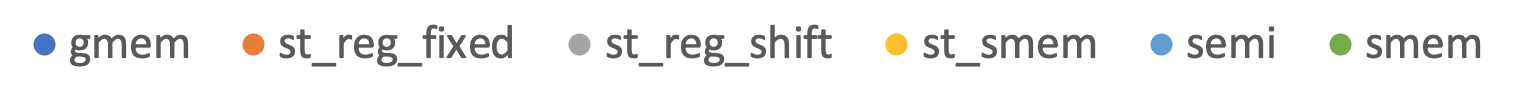}
         \label{fig:roofline-kernel-nameplate}
     \end{subfigure}
        \caption{Roofline Performance on V100}
        \label{fig:roofline}
\end{figure*}

Table \ref{table:performance-characterstics} presents the performance characteristics
    of our implementations on the \texttt{V100}.
Figure \ref{fig:roofline} visualizes these performance characteristics
    using roofline performance model,
    where Subfigures \ref{fig:roofline-device-l2} and \ref{fig:roofline-device-dram} 
        showing the rooflines for L2 and DRAM, respectively,
        and Subfigures \ref{fig:roofline-kernel-l2} and \ref{fig:roofline-kernel-dram} 
        are zoomed-in views the roofline kernel characteristics.
The y-axes of these figures represent performance 
    and x-axes show arithmetic intensity.
The dots in each group of the implementations are categorized with the same color
    and their coordinates can be found in Table \ref{table:performance-characterstics}.

From our results, we offer the following observations.

\bigskip
\subsubsection*{3D Blocking using Global Memory}\hfill\\
The simplest implementation \texttt{gmem\_8x8x8} using only the global memory
    yields the best performance on V100.
With L1 data cache and shared memory
    combined into a single unified memory block on the V100~\cite{v100-arch-whitepaper},
    we have a much larger data cache available on the V100 than on   previous generations of GPUs.
Therefore, when retrieving data from global memory
    with a good access pattern,
    we can achieve very good performance.

Comparing the performance of the 3D kernel across GPU generations,
    we notice its poor performance portability.
It is one of the slowest implementations on P100 and the NVS510.

We tried several variants of the global memory implementation
    that differ each others in terms of block size,
    from smaller to larger, including
    \texttt{gmem\_4x4x4},
    \texttt{gmem\_8x8x4},
    \texttt{gmem\_8x8x8},
    \texttt{gmem\_16x16x4},
    and \texttt{gmem\_32x32x1}.
Our results show that, \texttt{gmem\_8x8x8} is the best among them.
We need to load all halos before performing stencil computations.
    For blocks  smaller than \texttt{gmem\_8x8x8},
    such as \texttt{gmem\_4x4x4} and \texttt{gmem\_8x8x4},
    the halo size for our 25-point stencil dominates the actual data points.
Therefore, more time is spent on loading halos than points for the volume to be computed, which hurts performance.
In addition, smaller block sizes also result in larger GPU grid size
    as we can see from Table \ref{table:kernel-characterstics-v100},
    which means more kernel launches. For the smaller blocks, the additional overheads slow the overall execution.
On the other hand, we also see performance degradation
    for larger blocks, \texttt{gmem\_16x16x4} and \texttt{gmem\_32x32x1}.
Their larger block size results in a smaller grid size.
However,
    both have low theoretical and achieved occupancy.
Table \ref{table:performance-characterstics} shows that the 3D kernels using larger blocks, especially \texttt{gmem\_32x32x1}, incur more L2 cache misses, which increases the number of high-latency loads from  global memory.

In summary, the
    global memory implementations are the simplest to program
    and need very little performance tuning.
With the right tile shape and using a good global memory access pattern,
    on late-model GPU architectures, such as V100,
    one can achieve amazingly good performance with little effort.
From a software engineering perspective, 
    these implementations are easy to understand
    and have a low maintenance cost.

\bigskip
\subsubsection*{Shared memory}\hfill\\

Table \ref{table:time-measurement} shows that using shared memory can boost performance.
The yield performance gain is more significant on older generation GPUs, such as P100 and NVS510, which is consistent with results in previous research.

Recall that \texttt{smem\_u} is a high-order stencil
        while \texttt{smem\_eta\_1} and \texttt{smem\_eta\_3} are not.
From Table \ref{table:time-measurement},
we saw \texttt{smem\_u} runs faster than \texttt{smem\_eta\_1} and \texttt{smem\_eta\_3} on V100, but oppositely, it is slower on P100 and NVS510.
We attribute this conflicting results to the architectural changes in V100, 
where it combines the L1 data cache with shared memory.
As discussed previously, on V100, with good access patterns for global memory,
    one can achieve great performance with little effort.
The overhead of using shared memory on V100 in 3D blocking erases this gain.
In contrast, older generation GPUs do not have this new feature,
    so shared memory provides more performance benefits than its overhead.
On older architectures,
with high-order stencils, such as \texttt{smem\_u},
    because we load larger-size blocks into shared memory
    than low-order ones, such as \texttt{smem\_eta\_1} and \texttt{smem\_eta\_3},
    we see better performance.

For high-order stencils, which have a large halo size,
    it is not hard to reach the shared memory limit.
While shared memory improves performance,
    the hardware limitation on shared memory limits the potential of holding all data on shared memory for high-order stencils which use large blocks.

\bigskip
\subsubsection*{Semi-stencil}\hfill\\

Thread synchronizations on GPUs are very expensive,
    and our evaluations also prove so with our semi-stencil implementation.
Because of the need of storing and loading partial results,
    thread synchronizations are necessary
        to ensure the completeness of the required computation
        before it can proceed to the next.
As GPU runs threads concurrently in warps,
    we must introduce proper barriers
        to prevent data from being corrupted.
Our choice of using 3D blocking requires thread synchronizations 
    on all three dimensions,
    which exacerbates the problem.
HPCToolkit also backs our reasoning
    with its second most significant bottleneck
    being the thread synchronization (\texttt{STL\_SYNC}).
    
Nevertheless, the methodology behind semi-stencil algorithm is still valid.
Thus, to avoid the excess use of thread synchronizations,
    we will investigate into using double buffering.
As well, we will explore using 2.5D blocking instead of 3D blocking in our future work.

\bigskip
\subsubsection*{Code Shape for 2.5D-Blockings}\hfill\\

For implementations using 2.5D-blocking,
    we observe
    the larger the 2D plane, the better the performance.
There are two main reasons for this.
First, a larger 2D plane means a higher degree of concurrency.
Second, with a larger plane,
    the percentage of halo points fetched into shared memory is smaller,
    which speeds up the overall performance.

In addition, our results show that \texttt{st\_reg\_shft\_32x16} runs faster than \texttt{st\_reg\_shft\_16x32}.
From Table \ref{table:performance-characterstics},
    we see more L2 transactions with \texttt{st\_reg\_shft\_16x32},
    which in turn harms performance.
Therefore, one should cut the plane so that the x-dimension of the GPU's thread block assigned
to the innermost dimension 
has a relatively larger size.

\bigskip
\subsubsection*{Register Footprint in 2.5D-Blockings}\hfill\\
When we evaluate \texttt{st\_reg\_shft\_}$*$ implementations,
    the variants with 2D plane size of $1024$,
    namely
    \texttt{st\_reg\_shft\_16x64},
    \texttt{st\_reg\_shft\_32x32},
    and \texttt{st\_reg\_shft\_64x16},
    show poor performance on V100.
The performance degradation is caused by register spilling.
The maximum registers in a blockthread is $64*1024=65536$.
Because we have $1024$ threads for these implementations,
    we can only have maximum $64$ registers for each thread.
If we do not explicitly specify the register count to \texttt{nvcc},
    it assigns $80$ and $96$ registers
        to the PML and inner kernels, respectively.
Running the generated binaries for these register footprints yields incorrect results. To avoid this problem, we use compiler flag \texttt{-maxrregcount=64}
    to limit the maximum register usage per thread.
Unfortunately $64$ registers are 
not enough to hold all of the variables at the same time,
causing register spilling.
The register shifting approach exacerbates  register spilling 
    due to its high frequency of register access.

However, although register spilling  happens to the register shifting kernels,
for the \texttt{st\_fixed\_reg\_32x32} kernel, 
    we don't see a performance degradation because the code uses fixed registers with loop unrolling.
Because the registers are fixed,
    the frequency of register data movement
        is smaller than for kernels using the register shifting approach.
This allows the performance impacted by  register spilling
    to be hidden by other thread activities.
    
\bigskip
\subsubsection*{GPU Warp Occupancies}\hfill\\
Table \ref{table:kernel-characterstics-v100} shows
    implementations using 2.5D blocking 
    in general have better
    theoretical and achieved occupancies
    than the ones using 3D blocking.

\bigskip
\subsubsection*{Performance Portability}\hfill\\
The best performing implementations on P100 and NVS510
    come from 2.5D approaches.
Although they are not  the best kernels on V100,
    they are still in the fastest tier.
Thus, if performance portability is a concern,
    implementations using 2.5D blocking,
    such as \texttt{st\_reg\_fixed\_32x32},
    would be preferred.

\bigskip
\subsubsection*{Gaps to the Roofline Ceilings}\hfill\\
Our interpretation of the performance gaps are twofold:

First, 
although our current implementations realize a good performance 
    for high-order stencils with boundary conditions,
    we see room for further performance tuning.
We could improve the arithmetic intensities
    by designing new GPU code shapes, e.g., by employing the semi-stencil algorithm~\cite{semi1,semi2}, which reduces data movement for high-order stencils,
        or employing time-skewing to  increase data reuse. 
While all of our  implementations were manually written in CUDA,
    we could develop a new DSL approach or build a framework that
    enables us to explore more sophisticated approaches.

Second,
    ERT uses simple micro-benchmarks to profile the machines.
In contrast, the
    acoustic isotropic model not only uses  high-order stencils
        with complex boundary conditions,
    but also contains complicated logic with multiple statements.
Thus, while the roofline ceilings provide us a guide, the logic of our complex kernels makes it difficult to hit the roofline limit. 

\section{Conclusions and Future Work}
\label{sec:concl}

In this paper,
we evaluated the performance of high-order stencils with boundary conditions.
    We reviewed the existing techniques for implementing 3D stencil computations and evaluated most of them suitable for high-order stencils, with the notable exception of time-skewing algorithms.
We implemented multiple versions of different approaches
using known techniques and combinations of them,
modifying approaches as needed to  better accommodate high-order stencils with boundary conditions.
We evaluated our implementations
    with several tools on multiple platforms,
     quantitatively comparing stencils from various perspectives,
    and presented our findings and observations.

We  began our evaluation by computing 25-point stencil algorithms over the entire data domain in a single kernel launch. Inefficiency caused by branch divergence led us to apply domain decomposition and compute the boundaries separate from the center region. While this improved efficiency, having the regions separate impedes our ability to apply time skewing along the streaming dimension for the 2.5D algorithm.
We plan to reintegrate  boundary computations with the inner region to enable us to evaluate 3.5D algorithms on high-order stencils by adding time skewing to the streaming dimension.
In addition, we would like to explore whether applying the semi-stencil algorithm~\cite{semi1} along the streaming dimension to reduce  the memory hierarchy footprint of a block's stencil calculations and increase the arithmetic intensity of high-order kernels.
We plan to experiment with the range of kernels on the NVIDIA A100 GPU to assess performance portability.
In addition to NVIDIA GPUs, we plan to expand the scope of our evaluation to explore the performance of various high-order stencil implementations on leading-edge GPUs from other vendors, as soon as we can gain access to them and results on them are not embargoed.
    
We recognize that 
 implementing, tuning, maintaining, and porting high-performance GPU kernels for high-order stencils is quite difficult. For the long term, we believe that a high-level representation of stencil computations in conjunction with powerful compiler technology is arguably the best  strategy to improve   development productivity and performance portability while also lowering maintenance costs by reducing complexity. However, a concern for DSL users is the long-term viability of their code. We are hopeful that adding DSL technology to the LLVM compiler framework will provide a path forward that will address this concern.

\section*{Acknowledgment}

This work was supported in part by a contract from Total E\&P Research \& Technology USA, LLC.
We thank Keren Zhou from Rice University for reviewing the drafts of this paper and helping us use his emerging GPU Performance Advisor tool, which
 offered insights for tuning some of the kernels we studied.

\end{document}